\documentstyle[12pt]{article}

\def\hybrid{\topmargin -20pt    \oddsidemargin 0pt
        \headheight 0pt \headsep 0pt
        \textwidth 6.25in       % A4 paper
        \textheight 9.5in       % A4 paper
        \marginparwidth .875in
        \parskip 5pt plus 1pt   \jot = 1.5ex}

\hybrid

\catcode`\@=11

\def\marginnote#1{}
\newcount\hour
\newcount\minute
\newtoks\amorpm
\hour=\time\divide\hour by60
\minute=\time{\multiply\hour by60 \global\advance\minute by-\hour}
\edef\standardtime{{\ifnum\hour<12 \global\amorpm={am}%
        \else\global\amorpm={pm}\advance\hour by-12 \fi
        \ifnum\hour=0 \hour=12 \fi
        \number\hour:\ifnum\minute<10 0\fi\number\minute\the\amorpm}}
\edef\militarytime{\number\hour:\ifnum\minute<10 0\fi\number\minute}
\def\draftlabel#1{{\@bsphack\if@filesw {\let\thepage\relax
   \xdef\@gtempa{\write\@auxout{\string
      \newlabel{#1}{{\@currentlabel}{\thepage}}}}}\@gtempa
   \if@nobreak \ifvmode\nobreak\fi\fi\fi\@esphack}
        \gdef\@eqnlabel{#1}}
\def\@eqnlabel{}
\def\@vacuum{}
\def\draftmarginnote#1{\marginpar{\raggedright\scriptsize\tt#1}}

\def\draft{\oddsidemargin -.5truein
        \def\@oddfoot{\sl 2nd draft \hfil
        \rm\thepage\hfil\sl\today\quad\militarytime}
        \let\@evenfoot\@oddfoot \overfullrule 3pt
        \let\label=\draftlabel
        \let\marginnote=\draftmarginnote
   \def\@eqnnum{(\theequation)\rlap{\kern\marginparsep\tt\@eqnlabel}%
\global\let\@eqnlabel\@vacuum}  }

\def\preprint{\twocolumn\sloppy\flushbottom\parindent 2em
        \leftmargini 2em\leftmarginv .5em\leftmarginvi .5em
        \oddsidemargin -.5in    \evensidemargin -.5in
        \columnsep .4in \footheight 0pt
        \textwidth 10.in        \topmargin  -.4in
        \headheight 12pt \topskip .4in
        \textheight 6.9in \footskip 0pt
        \def\@oddhead{\thepage\hfil\addtocounter{page}{1}\thepage}
        \let\@evenhead\@oddhead \def\@oddfoot{} \def\@evenfoot{} }

\def\numberbysection{\@addtoreset{equation}{section}
        \def\theequation{\thesection.\arabic{equation}}}

\def\underline#1{\relax\ifmmode\@@underline#1\else
        $\@@underline{\hbox{#1}}$\relax\fi}

\newskip\humongous \humongous=0pt plus 1000pt minus 1000pt

\newif\ifdtup

\relax

\begin{document}

%\draft

\vspace{.4in}

\begin{center} \Large\bf 
Dynamical r-matrices\\ 
and the chiral WZNW phase 
space\footnote{Talk given at XXIII International Colloquium on Group
    Theoretical Methods in Physics, July 31 - August 5, 2000, Dubna, Russia.}

\end{center}

\vspace{.1in}

\begin{center}
L.~Feh\'er

\vspace{0.2in}       

{\em Institute for Theoretical Physics,
         University of Szeged} \\
       {\em  H-6726 Szeged, Tisza Lajos krt 84-86, Hungary }\\
{\em e-mail: lfeher@sol.cc.u-szeged.hu}\\

\end{center}

\vspace{.2in}

\begin{center} \bf Abstract
\end{center}

{\parindent=25pt
\narrower\smallskip\noindent
The dynamical generalization of the classical Yang-Baxter 
equation that governs the
possible Poisson structures on the space of chiral WZNW fields with
generic monodromy is reviewed.
It is explained that for particular choices of the chiral WZNW Poisson brackets
this  equation reduces to the CDYB equation recently studied
by Etingof--Varchenko and others.
Interesting dynamical r-matrices are obtained for generic monodromy  
as well as by imposing Dirac constraints on the monodromy. 
}
  
\vfill\eject

%%MACROS USED IN THIS PAPER BY L. FEHER %
                                        %                                      
\def\tr{{\rm Tr\,}}                     %
\def\Ad{{\rm Ad\,}}                     % 
\def\ad{{\rm ad\, }}                    %
\def\G{{\cal G}}                        %
\def\M{{\cal M}}                        %
\def\H{{\cal H}}                        %
\def\D{{\cal D}}                        %
\def\L{{\cal L}}                        %
\def\cR{{\cal R}}                       %
\def\R{{\mathbf R}}                     %
                                        %
%%%%%%%%%%%%%%%%%%%%%%%%%%%%%%%%%%%%%%%%%

\section{Introduction}

The classical and quantum Yang-Baxter equations
occupy a central position in the modern theory of integrable systems.
Recently dynamical generalizations 
of these structures attracted considerable  attention. 
Of particular interest for us is the classical dynamical 
Yang-Baxter (`CDYB') equation given by
\begin{equation}
\big[\hat r_{12}(\omega),\hat r_{23}(\omega)\big]
+H_1^i \frac{\partial}{\partial \omega^i} \hat r_{23}(\omega)
+ \hbox{cycl. perm.}=C \hat f,
\label{CDYB}\end{equation}
where the variable $\omega$ lies in a Cartan subalgebra
$\H$ of a simple Lie algebra $\G$, $\{ H^i\}$ is a basis of $\H$,
$C$ is some constant, and $\hat f$ is the canonical $\G$-invariant
element in $\G^{3\wedge}$.
It is usually assumed that $\hat r(\omega)$  is $\H$-invariant
and its symmetric part is proportional
to the `tensor Casimir'. 
The CDYB equation is the classical limit of the
Gervais-Neveu-Felder equation
\begin{equation}
R_{12}(\omega + \hbar H_3) R_{13}(\omega) R_{23}(\omega + \hbar H_1)
=R_{23}(\omega) R_{13}(\omega + \hbar H_2) R_{23}(\omega).
\end{equation}
These equations govern the classical and quantum exchange
algebras of the chiral Bloch waves in the conformal Toda
and WZNW field theories on the cylinder \cite{GN,CG,BDF}.
They also appear in the description of the conformal blocks
of the WZNW model on the torus \cite{Feld} 
and in the study of Calogero-Moser models \cite{ABB}. 
The solutions of these equations and the underlying 
 abstract algebraic structures, the so called dynamical 
Poisson-Lie (PL) groupoids and dynamical quantum groups, 
have been studied recently in details by Etingof and Varchenko.
See the review \cite{ES} and references therein, 
where further applications are described, too.
In the paper \cite{EV} generalizations
of the CDYB equation  were introduced,
which are obtained from (\ref{CDYB})
by replacing the Cartan subalgebra with 
an arbitrary  subalgebra of $\G$.
We here call this generalization 
the $\H$-CDYB equation allowing $\H\subseteq \G$ to be any 
subalgebra.

We have recently investigated the 
possible chiral extensions of the WZNW 
phase space and found that a new generalization 
of the CDYB equation naturally arises in this context \cite{BFP}.
This equation will be called the $G$-CDYB equation,
since its dynamical variable lies in the group $G$ associated with $\G$.
The $G$-CDYB equation encodes
the most general PBs of the chiral WZNW fields with generic monodromy.
Any solution of this equations defines also a PL groupoid.
Under some special circumstances, when the 
natural gauge transformations act on the chiral WZNW phase space 
as a classical $\G$-symmetry, our $G$-CDYB reduces
 to $\G$-CDYB (i.e. $\H$-CDYB for $\H=\G$).
Then Dirac reductions of the chiral WZNW phase space result in
dynamical r-matrices that solve $\H$-CDYB 
for self-dual subalgebras $\H\subset \G$.
For instance, we recover in this way the fundamental 
solution of the original CDYB equation that was first
obtained in \cite{BDF} by a different method.

It will be illustrated by this report that the chiral WZNW phase space 
serves as an effective
source of dynamical r-matrices.
The quantization of these r-matrices and their
associated Poisson-Lie groupoids should contribute to a better understanding
of  the quantum group properties of the 
WZNW model, but this issue is not yet properly understood.

\section{$G$-CDYB and PL groupoids from chiral WZNW }

The WZNW model \cite{Wi} as a classical field theory on the cylinder
can be defined for any (real or complex) Lie group $G$
whose Lie algebra $\G$ carries an invariant, non-degenerate 
bilinear form $\langle\ ,\ \rangle$.
The scalar product is proportional to ${\rm tr}(XY)$ if $\G$ 
is a simple Lie algebra,
and to ease the notations we shall denote 
$\langle X,Y\rangle$ ($\forall X,Y\in \G$)  as 
$\tr(XY)$ in general.
The solution of the classical field equation for the $G$-valued
WZNW field, which is $2\pi$-periodic in the space variable,
turns out to be the product of left- and right-moving 
factors.
The chiral WZNW fields are quasi-periodic, i.e., are elements in
\begin{equation}
\M_G:= \{  g \in C^\infty({\R}, G)  \,\vert\,
g(x + 2\pi) = g(x) M \quad  M\in G\}.
\label{MG}\end{equation}
Since the chiral factors of the full WZNW field
are determined only up to a gauge freedom, the symplectic structure
of the WZNW model does not yield a unique Poisson bracket (PB) 
on  $\M_G$. 
In fact, as explained in \cite{BFP+},
$\M_G$ is  equipped  canonically only with a quasi-Poisson 
structure in the sense of \cite{AKS}. 
To describe the system in terms of  
genuine PBs for the chiral fields 
and an associated chiral symplectic form  \cite{G}, 
in general one needs to restrict oneself to a submanifold
of $\M_G$, where the monodromy matrix $M$ belongs to some 
submanifold $\check G\subset G$.
A condition on $\check G$ 
is that the canonical closed 3-form of $G=\{ M\}$, given by 
$\chi =\frac{1}{6} \tr\left( M^{-1}dM \wedge M^{-1}dM\wedge M^{-1}dM\right)$,
must become exact upon restriction to $\check G\subset G$.
One may then choose a 2-form $\rho$ on $\check G$ for which
$d \rho = \chi_{\check G}$,
where $\chi_{\check G}$ is the restriction of $\chi$ to $\check G$.
For any such $\rho$,  one can 
define a closed $2$-form $\Omega^{\rho}$ on  
\begin{equation}
\M_{\check G}:= 
\{ g  \in C^\infty({\R}, G)\,\vert\, 
 g(x + 2\pi) = g(x) M \quad  M\in \check G\}
\label{McheckG}\end{equation}
by the following formula:
\begin{equation}
\frac{1}{\kappa} \Omega^\rho=
- \frac{1}{2}\int_0^{2\pi}dx\,
 \tr\left(g^{-1}dg \right) \wedge \left(g^{-1}dg \right)'
-\frac{1}{2} \tr \left( (g^{-1} dg)(0)
\wedge dM 
{\scriptstyle\,} M^{-1}\right) + \rho(M),
\label{Omegarho}\end{equation}
where $\kappa$ is a constant.
If a further condition is satisfied, which we shall state below, 
then $\Omega^{\rho}$ is (weakly) non-degenerate, and thus it can be
inverted to define PBs on a set of `admissible' 
functions of the chiral WZNW field.
The derivation of $\Omega^\rho$ from the
symplectic structure of the full WZNW model is due to 
Gawedzki \cite{G}.

One may ensure the exactness of $\chi_{\check G}$ 
by choosing $\check G\subset G$ to be a topologically trivial 
open submanifold. 
In this case, the following
description of the chiral PBs
was obtained in \cite{BFP} by extending the results of \cite{FG}.
In fact,  
the PBs of all admissible functions \cite{BFP} 
are encoded by the `distribution valued' PBs
of the matrix elements of $g(x)$, which have the form
\begin{equation}
\Big\{g(x)\stackrel{\otimes}{,} g(y)\Big\}
=\frac{1}{\kappa}\Big(g(x)\otimes
g(y)\Big)\Big(
\frac{1}{2} {\hat I} \,{\rm sign}\,(y-x) +\hat r(M)  
\Big), \quad 0< x,y<2\pi.
\label{gPB}\end{equation}  
Here the interesting object is the `exchange r-matrix'
$\hat r(M)= r^{ab}(M) T_a \otimes T_b \in \G\wedge \G$;
$\hat I=T_a \otimes T^a$ where $\{ T_a\}$ and $\{T^a\}$ 
denote dual bases of $\G$, $\tr(T_a T^b)=\delta_a^b$, and 
summation over coinciding indices is understood.
The Jacobi identity of the PB is equivalent 
to a dynamical generalization of the CYB equation, 
which we call the $G$-CDYB equation.
To write it down, on functions $\psi$ on $G$ we introduce the
derivations
$\D_a^{\pm} = \cR_a \pm \L_a$ by
\begin{equation}
({\cal R}_a \psi)(M):= \frac{d}{d t}
 \psi(Me^{t T_a} )\Big\vert_{t=0},
\qquad 
({\cal L}_a \psi)(M):= \frac{d}{d t}
 \psi(e^{t T_a} M)\Big\vert_{t=0}.
\end{equation}
The $G$-CDYB equation \cite{BFP} reads as 
\begin{equation}
\Big[\hat r_{12}(M),\hat r_{23}(M)\Big]
+T_1^a \Big( \frac{1}{2} \D_a^+ + 
r_a^{{\phantom{a}}b}(M) \D_b^-\Big) \hat r_{23}(M)
+ \hbox{cycl. perm.}=
-\frac{1}{4} \hat f,
\label{GCDYB}\end{equation}
where
$\hat f:= f_{a b}^{\phantom{ab}c} T^a \otimes T^b \otimes T_c$ with 
$[T_a, T_b]=f_{ab}^{\phantom{ab}c}T_c$ 
and the cyclic permutation is over the three tensorial factors
with  
$\hat r_{23} = r^{ab}  (1\otimes T_a\otimes T_b)$,
$T^a_1= T^a \otimes 1 \otimes 1$ and so on.
This equation becomes the modified classical YB 
equation if $\hat r$ is an $M$-independent constant,
and at the same time it is a generalization of the CDYB equation (\ref{CDYB}). 

The exchange r-matrix that results from the inversion
of a symplectic form in (\ref{Omegarho}) automatically satisfies (\ref{GCDYB}).
To describe its dependence on the 2-form $\rho$,
expand $\rho$ as $\rho(M)= \frac{1}{2}\,q^{ab}(M){\tr}
\big(T_a M^{-1}dM\big)\wedge{\tr}\big(T_b M^{-1}dM\big)$,
where $q^{ab}=-q^{ba}$.
Denote by $q(M)$ and $r(M)$ the linear operators on
$\G$ whose matrices are  $q^{ab}(M)$ and $r^{ab}(M)$,
respectively.
Introduce also the operators 
$q_\pm(M):= q(M) \pm \frac{1}{2}I$ and 
$r_\pm(M):= r(M) \pm \frac{1}{2}I$, where
$I$ is the identity operator on $\G$.
It is proved in \cite{BFP} that the inversion of $\Omega^\rho$ leads
to (\ref{gPB}) with
\begin{equation}
r_-(M)= - q_-(M) \circ \left(q_-(M) - \Ad M\circ q_+(M)\right)^{-1}.
\label{r-rho}\end{equation}
The condition on the pair $(\check G, \rho)$ that guarantees
the non-degeneracy of the 2-form $\Omega^\rho$  is that 
 $(q_-(M) - \Ad M\circ q_+(M))\in {\rm End}(\G)$ must be an invertible
operator for any $M\in \check G$.
This can be ensured by restricting $M$ to be near enough to $e\in G$.

Any solution of (\ref{GCDYB}) on some domain $\check G$ gives
rise to a PB (\ref{gPB}) on $\M_{\check G}$ and any such PB 
implies that $J:= \kappa g' g^{-1}$ satisfies 
the standard current algebra PBs 
and $g(x)$ is a primary field with respect to the current algebra. 
The exchange r-matrix drops out from the PBs 
with any function of the current $J$, and thus
it encodes the `non-current-algebraic' aspects of the infinite
dimensional chiral WZNW phase space. 
Remarkably, the exchange r-matrix also defines the PBs of an associated 
 Poisson-Lie groupoid, as described below.

Let $\hat r$ be a solution of (\ref{GCDYB}) on $\check G$
and define $\G\otimes \G$-valued functions on $\check G$ by
\begin{equation}
\hat \Theta(M) = \hat r_+(M)  -M_2^{-1} \hat r_-(M) M_2\,,\qquad
\hat \Delta(M) = \hat \Theta(M) - M_1^{-1} \hat \Theta(M) M_1 
\label{Delta}\end{equation} 
with $M_1= M\otimes 1$, $M_2= 1\otimes M$. 
Then introduce on the manifold $P$ given by
\begin{equation}
P:= \check G \times G\times \check G:= \{ (M^F, g, M^I)\}
\end{equation}
a PB $\{\ ,\ \}_P$ by the following formulas:
\begin{eqnarray}
&&
\kappa \{ g_1, g_2\}_P = g_1 g_2 \hat r(M^I) - 
\hat r(M^F) g_1 g_2 
\nonumber\\
&& \kappa \{ g_1, M^I_2\}_P = g_1 M_2^I \hat \Theta(M^I)
\nonumber\\
&& \kappa \{ g_1, M_2^F\}_P = M_2^F \hat\Theta(M^F) g_1
\nonumber\\
&&\kappa \{ M^I_1, M^I_2\}_P = M^I_1 M^I_2 \hat\Delta(M^I)
\nonumber\\
&&\kappa \{ M^F_1, M^F_2\}_P = - M^F_1 M^F_2 \hat\Delta(M^F)
\nonumber\\
&&\kappa \{ M^I_1, M^F_2\}_P =0.
\label{PPB}
\end{eqnarray}
$P$ is an example of the simplest sort of groupoids \cite{MacK}:
the base is $\check G$, 
the source and target projections operate as
$s: (M^F, g, M^I) \mapsto M^I$ and
$t: (M^F, g, M^I)\mapsto M^F$, 
and the partial multiplication is defined by  
$(M^F, g, M^I) (\bar M^F, \bar g, \bar M^I):= (M^F, g\bar g, \bar M^I)$
for
$M^I=\bar M^F$.
$P$ is a Poisson-Lie groupoid in the sense of \cite{We}.
This means that the graph of the partial multiplication,
i.e.~the subset of 
\begin{equation}
P\times P\times P =
\{ (M^F, g, M^I)\} \times \{ (\bar M^F, \bar g, \bar M^I)\}
\times \{ (\hat M^F, \hat g, \hat M^I)\} 
\end{equation}
defined by the constraints
$M^I=\bar M^F$,
$\hat M^F= M^F$,
$\hat M^I=\bar M^I$, 
$\hat g= g\bar g$,
is a coisotropic submanifold of  
$P\times P\times P^-$, where $P^-$ denotes the  manifold
$P$ endowed with the opposite of the PB on  $P$.
In other words, the graph  is defined   
by imposing  first class constraints on the Poisson space
$P\times P\times P^-$ equipped with the natural direct product PB.
This would actually hold for any choice of the structure functions
$\hat r(M)$, $\hat \Theta(M)$ and $\hat \Delta(M)$ in (\ref{PPB}),
and the choice (\ref{Delta}) in terms of a solution of (\ref{GCDYB})
guarantees the Jacobi identity for $\{\ ,\ \}_P$.

We have extracted a PL groupoid from  
any symplectic structure $\Omega^\rho$  on the chiral 
WZNW phase space.  
If the exchange r-matrix   
is constant, then the PL groupoid $P$  
carries the same information 
as the group $G$ endowed with the corresponding Sklyanin bracket.
It is an open problem to study these PL
groupoids further in the general case,
to understand their quantization and relate them 
to the quantized (chiral) WZNW conformal field theory.

\section{$\G$-CDYB from $\G$-symmetry and $\H$-CDYB \\
 from Dirac reductions}

We next describe an interesting special case of the 
chiral WZNW symplectic structure $\Omega^\rho$, 
for which the corresponding exchange 
r-matrix becomes a solution of the $\G$-CDYB equation 
mentioned in the introduction, and then consider some Dirac reductions.

Let us suppose that $\check G$ is diffeomorphic to a domain 
$\check \G\subset \G$ by the exponential parametrization, whereby
 we write $\check G\ni M =e^{\omega}$ with $\omega\in \check \G$.
The chiral WZNW fields whose monodromy lies in $\check G$ 
can be parametrized as 
\begin{equation}
g(x) =\eta(x) e^{\bar\omega x},\qquad
\bar\omega:= \frac{\omega}{2\pi},\quad\eta\in \widetilde{G}.
\label{expar}\end{equation}
Here 
$\widetilde{G}=\{ \eta\in C^\infty(\R,G)\,\vert\, \eta(x+2\pi)=\eta(x)\,\}$
and equation (\ref{expar}) defines the identification
$\M_{\check G}= \widetilde{G} \times \check \G=\{ (\eta, \omega)\}$.
If we now choose the 2-form $\rho$ on $\check G \simeq \check \G$ to be
\begin{equation}
\rho_0(\omega)= -\frac{1}{2} \int_0^{2\pi} dx 
\tr\left(d\bar \omega \wedge de^{x\bar\omega} \, e^{-x\bar\omega}\right), 
\end{equation}
then in terms of the  variables $\eta$ and $\omega$ we find 
\begin{equation}
\frac{1}{\kappa} \Omega^{\rho_0}=
- \frac{1}{2 }\int_0^{2\pi}dx\,
 \tr\left(\eta^{-1}d\eta \right) \wedge \left(\eta^{-1}d\eta  \right)'
+d \int_0^{2\pi} dx 
\tr\left(\bar\omega \eta^{-1} d\eta \right).
\label{Omegaexp}\end{equation}
$\Omega^{\rho_0}$  is 
invariant under the natural 
action
of the group $G$ 
on $\M_{\check G}$ given by
\begin{equation}
G\ni h: g(x) \mapsto g(x)h^{-1} 
\quad\mbox{i.e.}\quad 
\eta(x) \mapsto \eta(x) h^{-1},
\quad
\omega \mapsto h \omega h^{-1}.
\label{Gsym}\end{equation}
Note that $\check \G$ is assumed to be  
invariant under the action of $G$, otherwise one has to  
consider the corresponding $\G$-action.
Since $\Omega^{\rho_0}$ is symplectic, 
which may be ensured by taking $\check \G$ to be a  
neighbourhood of $0$, the $\G$-symmetry obtained from (\ref{Gsym})  
is generated by a classical momentum map.
The value of this $\G\simeq \G^*$ valued momentum map
is proportional to $\omega$.
In fact, we can calculate that 
the PB $\{\ ,\ \}_0$ corresponding to $\Omega^{\rho_0}$ gives  
\begin{equation}
\{ g(x), \omega_a \}_0 = \frac{1}{\kappa}g(x) T_a,
\qquad\qquad
\{ \omega_a, \omega_b\}_0 =-\frac{1}{\kappa} 
f_{ab}^{\phantom{ab}c} \omega_c.
\label{Ymoment}\end{equation} 
Moreover, we have  
\begin{equation} 
\Big\{g(x)\stackrel{\otimes}{,} g(y)\Big\}_0
=\frac{1}{\kappa}\Big(g(x)\otimes
g(y)\Big)\Big(
{1\over 2} {\hat I} \,{\rm sign}\,(y-x) +\hat r^0(\omega)  
\Big), \quad 0< x,y<2\pi,
\label{g0PB}\end{equation}   
where $\hat r^0(\omega)$ denotes the exchange r-matrix 
associated with $\rho_0(\omega)$ by
(\ref{r-rho}). 
Now the Jacobi identity of the PB for the functions $\omega_a, g(x), g(y)$
$(x\neq y)$ 
and the relations in (\ref{Ymoment}) imply that $\hat r^0$ is
a $\G$-equivariant function on $\check \G$:
\begin{equation}
{d\over dt} \hat r^0 (e^{tT} \omega e^{-tT})\vert_{t=0} = 
[ T\otimes 1 + 1\otimes T, \hat r^0(\omega)]
\qquad
\forall T\in \G. 
\label{Gequivar}\end{equation}
In the present case, 
the Jacobi identity for 3 evaluation functions $g(x_i)$
($x_i \neq x_j$) gives
a simplified version of the $G$-CDYB equation (\ref{GCDYB}).
Namely, the Jacobi identity and (\ref{Ymoment}) imply that
\begin{equation}
\big[\hat r^0_{12}(\omega),\hat r^0_{23}(\omega)\big]
+T_1^a \frac{\partial}{\partial \omega^a} \hat r_{23}^0(\omega)
+ \hbox{cycl. perm.}=
-\frac{1}{4} \hat f
\qquad (\check \G\ni \omega=\omega^a T_a).
\label{cGCDYB}\end{equation}
This is nothing but the $\G$-CDYB equation mentioned in the introduction.
We stress that this equation follows as a consequence of 
the Jacobi identity of the PBs (\ref{g0PB}) and (\ref{Ymoment}).

We can now determine $r^0(\omega)$ explicitly from (\ref{r-rho}) and 
thereby find a solution of (\ref{cGCDYB}).
The result \cite{BFP} is given by 
$r^0(\omega) = f_0(\ad \omega)$  with $f_0$ being the power series
expansion of the complex analytic function 
\begin{equation}
f_0(z) = \frac{1}{2}\coth \frac{z}{2} - \frac{1}{z}
\label{f0}\end{equation}
around $z=0$.
This solution of (\ref{cGCDYB}) was found in a  
different context \cite{AM}, too. 

In \cite{EV} the CDYB equation (\ref{CDYB}) was generalized  
by allowing the dynamical variable to belong to
the dual of an arbitrary subalgebra $\H\subset \G$.
Next we explain that if $\H\subset \G$
is a `self-dual' subalgebra, then some solutions of the
$\H$-CDYB equation  
arise from the solutions of (\ref{Gequivar}) and (\ref{cGCDYB}) 
upon applying Dirac reduction to the associated PB on 
$\M_{\check G}$.   

We now start by considering a PB of the form (\ref{g0PB}) 
on $\M_{\check G}$ and also suppose that (\ref{Ymoment}) holds
where $\omega=\log M$ varies in a domain $\check\G  \subset \G$.
As we have seen, then the exchange r-matrix  $\hat r^0(\omega)\in \G\wedge \G$
satisfies (\ref{Gequivar}) and (\ref{cGCDYB}).
We choose a Lie subalgebra $\H\subset \G$ and assume 
that the restriction of the scalar product of $\G$ remains
non-degenerate on $\H$, 
which means that $\H$ is `self-dual'.
We have the linear direct sum decomposition 
$\G= \H + \H^\perp$
and can introduce an adapted basis of $\G$ 
in the form
$\{ T_a\} = \{ H_i\} \cup \{ E_\alpha\}$,
$H_i \in \H$, $E_\alpha \in \H^\perp$
with dual basis $\{ T^a\} = \{ H^i\} \cup \{ E^\alpha\}$.
(The notation is motivated by the
`principal example' for which
$\H$ is a Cartan subalgebra of a simple Lie algebra.)
Correspondingly, we can write
\begin{equation}
\omega= \omega_\H + \omega_{\H^\perp} = \omega^i H_i + \omega^\alpha E_\alpha.
\end{equation}  
We wish to impose the Dirac constraint
$\omega_{\H^\perp}=0$ on the PB on $\M_{\check G}$.
To calculate the resulting Dirac bracket,
we need to invert the matrix 
${\cal C}^{\alpha\beta}(\omega_\H) := 
\{ \omega^\alpha, \omega^\beta\}_0\vert_{\omega_{\H^\perp}=0}$. 
This is identified from (\ref{Ymoment}) 
as  the matrix of the linear operator 
${\cal C}(\omega_\H): \H^\perp \rightarrow \H^\perp$ 
that equals the restriction of a multiple of $\ad \omega_\H$ to $\H^\perp$,
\begin{equation}
{\cal C}(\omega_\H):= \frac{1}{\kappa} \ad \omega_\H \vert_{\H^\perp}.
\end{equation}
Thus we also have to 
restrict ourselves to a submanifold
of $\M_{\check G}$ where ${\cal C}(\omega_\H)$ is invertible.
We define the domain $\check\H\subset \check\G$ to contain the
$\H$-projection of those elements $\omega\in \check\G$ for which the operator 
${\cal C}^{-1}(\omega_\H)$ exists. 
Then we can compute the Dirac bracket on the constrained manifold
$\M_{\check\H}:= \{ g \in \M_{\check G}\,\vert\,  \log M \in \check\H\,\}$
by using the standard formula 
$\{ F_1, F_2\}_0^* = \{ F_1, F_2\}_0 - 
\{ F_1, \omega^\alpha\}_0{\cal C}^{-1}_{\alpha\beta}(\omega)
\{\omega^\beta, F_2\}_0$.
From (\ref{Ymoment}) we obtain
\begin{equation}
\{ g(x), \omega_i \}_0^* = {1\over \kappa} g(x) H_i,
\qquad
\{ \omega_i, \omega_j\}_0^* = -\frac{1}{\kappa} f_{ij}^{\phantom{ij}k} \omega_k
\qquad ([H_i, H_j]= f_{ij}^{\phantom{ij}k} H_k),
\label{Y*moment}\end{equation} 
where  $g(x+2\pi)= g(x)M$ with $\log M=\omega=\omega^i H_i\in \H$. 
Furthermore, (\ref{g0PB}) yields 
\begin{equation} 
\Big\{g(x)\stackrel{\otimes}{,} g(y)\Big\}_0^*
=\frac{1}{\kappa}\Big(g(x)\otimes
g(y)\Big)\Big(
\frac{1}{2} {\hat I} \,{\rm sign}\,(y-x) +\hat r^*(\omega)  
\Big), \quad 0< x,y<2\pi,
\label{g*PB}\end{equation}   
where
\begin{equation}
\hat r^*(\omega)= \hat r^0(\omega) + 
\frac{1}{\kappa}{\cal C}_{\alpha\beta}^{-1}(\omega) E^\alpha\otimes E^\beta.
\label{r*}\end{equation}
Now the point is that ---analogously to (\ref{Gequivar}) 
and (\ref{cGCDYB})--- 
the Jacobi identities of the
Dirac bracket imply that the function 
$\hat r^*: \check \H \rightarrow  \G\wedge \G$
given by (\ref{r*}) is $\H$-equivariant in the natural 
sense 
and satisfies the $\H$-CDYB equation
\begin{equation}
\big[\hat r^*_{12}(\omega),\hat r^*_{23}(\omega)\big]
+H_1^i \frac{\partial}{\partial \omega^i} \hat r_{23}^*(\omega)
+ \hbox{cycl. perm.}=
-\frac{1}{4} \hat f
\qquad ( \omega\in \check \H).
\label{HCDYB}\end{equation}

Examples may be  
obtained by taking $\H$ to be the grade zero subalgebra
in some integral gradation of 
$\G$ and taking $r^0(\omega)$ to be $f_0(\ad \omega)$.
Then the operator $r^*(\omega)$ ($\omega\in \check\H$)
associated with  (\ref{r*})  is found to be 
\begin{equation}
r^*(\omega)(X) =f_0(\ad \omega)(X)\quad \forall X\in \H,
\qquad
r^*(\omega)(Y)=
\frac{1}{2}\coth(\frac{1}{2}\ad\omega)(Y)\quad
\forall Y\in \H^\perp.
\end{equation}
We here use the Laurent series expansion 
of $\frac{1}{2}\coth(\frac{z}{2})$ in a punctured disc
around $z=0$,
and the $\frac{1}{z}$ term in the expansion corresponds to
the operator $(\ad\omega)^{-1}$ on $\H^\perp$.
In the special case of the principal gradation
of a simple Lie algebra $\G$,
for which $\H$ is a Cartan subalgebra,
this gives 
\begin{equation} 
\hat r^*(\omega)= \frac{1}{2} 
\sum_{\alpha\in \Phi} \coth\left(\frac{1}{2} \alpha(\omega)\right) 
E_\alpha \otimes E^\alpha,
\label{Hsol}\end{equation}
where $\Phi$ is the set of the roots and the $E_\alpha$ are 
corresponding root vectors. 
This solution of the CDYB equation
(\ref{CDYB}) was obtained  in \cite{BDF}
by determining the PBs of the chiral WZNW
Bloch waves with the aid  of a different method.
 
\section{Conclusion}

We reported on our recent results \cite{BFP}  
concerning the chiral WZNW phase
space by focusing on the dynamical generalizations of
the CYB equation that appear naturally in this context.
Not only several variants of the  CDYB
equation but also some of their most
interesting solutions were described.
The Dirac reduction of certain solutions of the $\G$-CDYB equation
to solutions   
of the $\H$-CDYB equation is treated in this report in a general manner
for the first time. 
Other aspects of the chiral WZNW phase space
that are not mentioned here for lack of space
have been also analysed in \cite{BFP,BFP+,BFPwak},
for instance we have found explicit solutions of the $G$-CDYB
equation that realize arbitrary PL symmetries.
The open problems that arose
from our investigation will hopefully be discussed in future publications.

\newpage
{\bf Acknowledgments.}
I wish to thank J. Balog and L. Palla for the collaboration
on which this report is based.
The work was supported in part by the Hungarian National
Science Fund (OTKA) under T025120,  T029802, T030099
and by the Ministry of Education under FKFP 0596/1999.


\begin{thebibliography}{99}

\bibitem{GN}
J.-L. Gervais and A. Neveu,  Nucl. Phys. B {\bf 238}, 125 (1984).
\bibitem{CG}
E. Cremmer and J.-L. Gervais, Commun. Math. Phys. {\bf 134},  619 (1990).
\bibitem{BDF}
J. Balog, L. D\c{a}browski and L. Feh\'{e}r, 
Phys. Lett. B {\bf 244}, 227 (1990). 
\bibitem{Feld} 
G. Felder,
Conformal field theory and integrable systems
associated to elliptic curves, %preprint {\tt hep-th/9407154}.
 pp.~1247-1255 in: {\em Proc. Int. Congr. Math. Z\"urich,   
1994} (Birkh\"auser, Z\"urich, 1994).
\bibitem{ABB} J. Avan, O. Babelon and E. Billey, 
Commun. Math. Phys. {\bf 178},  281 (1996).
\bibitem{ES} P. Etingof and O. Schiffmann,
Lectures on the dynamical Yang-Baxter equations,
preprint  {\tt math.QA/9908064}.
\bibitem{EV} P. Etingof and A. Varchenko, 
Commun. Math. Phys. {\bf 192}, 77 (1998).
\bibitem{BFP}
J. Balog, L. Feh\'er and L. Palla,
Nucl. Phys. B {\bf 568},  503 (2000).
\bibitem{Wi}
E. Witten, Commun. Math. Phys.  {\bf 92},   455 (1984).
\bibitem{BFP+}
J. Balog, L. Feh\'er and L. Palla,
The chiral WZNW phase space as a quasi-Poisson space,
preprint {\tt hep-th/0007045}. 
\bibitem{AKS}
A. Alekseev and Y. Kosmann-Schwarzbach, Manin pairs and moment maps,
preprint {\tt math.DG/9909176}. 
\bibitem{G}
K. Gaw\c{e}dzki, Commun. Math. Phys. {\bf 139},  201 (1991).
\bibitem{FG}
F. Falceto and K. Gaw\c{e}dzki, 
J. Geom. Phys. {\bf 11}, 251 (1993).
\bibitem{MacK}
K. Mackanzie, {\em Lie Groupoids and Lie Algebroids in Differential Geometry}
(Cambridge University Press, Cambridge,  1987). 
\bibitem{We}
A. Weinstein, J. Math. Soc. Japan {\bf 4},  705 (1988).
\bibitem{AM} 
A. Alekseev and E. Meinrenken,
Invent. Math. {\bf 139},  135 (2000).
\bibitem{BFPwak}
J. Balog, L. Feh\'er and L. Palla, 
J. Phys. A. {\bf 33}, 945 (2000).

\end{thebibliography}
\end{document}